\documentclass[fleqn,twoside]{article}
\usepackage{espcrc2}

\usepackage{graphicx}
\usepackage[figuresright]{rotating}

\newcommand{\AmS}{{\protect\the\textfont2
  A\kern-.1667em\lower.5ex\hbox{M}\kern-.125emS}}

\title{Progress in building an International Lattice Data Grid}

\author{A.C.~Irving\address{Theoretical Physics Division,    
         Department of Mathematical Sciences,
         University of Liverpool,
         Liverpool L69 3BX, UK},
	 R.D.~Kenway\address{School of Physics, 
	 University of Edinburgh, Edinburgh EH9~3JZ, UK},
	 C.M.~Maynard\addressmark and 
	 T.~Yoshi\'e\address{Center for Computational Physics, University of Tsukuba,
         Tsukuba, Ibaraki 305-8577, Japan}}
\begin{document}

\begin{abstract}
We report on progress in setting up the International Lattice Data
Grid.  We describe the aims and objectives of the ILDG, what has been
achieved during its first year of activity and invite feedback from
the community.
\vspace{1pc}
\end{abstract}

\maketitle
\section{INTRODUCTION}
\label{sec:intro}
Since the announcement of the International Lattice Data Grid (ILDG)
project at Lattice 2002~\cite{ILDG02} significant progress has been
made toward its initial objectives. The aim of the ILDG is to develop
grid-based technologies that enable simple and reliable exchange of
data between those international lattice research groups who choose to
use it. It is possible that the ILDG vision may develop into something
much more comprehensive than the simple publishing or exchange of
gauge configurations. This remains to be seen.  The initial proposal
emerged out of the QCDgrid project which is being developed by UKQCD
as a means of securely storing data and making it readily accessible
to its members irrespective of geographical location. The ILDG was
envisaged as a means of coordinating and promoting similar initiatives
by other international groups for their mutual benefit.

In the last year, there have been two (virtual) meetings
of ILDG participants from Australia, Europe, Japan and the USA.
In December 2002, two working groups were set up 
to drive forward the necessary technology: 
a Middleware Working Group (MWWG)~\cite{ref:MWWGmember}
and a Metadata Working Group (MDWG)~\cite{ref:MDWGmember}.

In the following sections, some of their achievements and
the issues exposed by their work are described. 
Broadly-speaking, the MDWG is charged with defining 
(via XML schema) the metadata required for grid storage
of lattice data. The MWWG is charged with defining standards
for the software interfaces of the middleware used to 
manipulate and manage the data contributed from local or national
grids and other storage systems. 
 
The initial work
of these groups was reviewed at a virtual meeting (via AccessGrid) 
in May 2003 and their medium and short terms goals revised.
The short term goals (i.e. by  LATTICE 2003) 
included 
presentation of an initial XML schema (QCDML) capable of describing
gauge configurations generated by a range of algorithms and actions
and demonstration of remote browsing of lattice data
collections in 3 continents.  

\section{MIDDLEWARE}
\label{sec:middleware}
First we review some general grid concepts and then report on how
these are being, or might be, applied in the context of ILDG. One
might well ask ``what {\em }is a grid and why do we want one?''.  The
grid is sometimes described as the next iteration in the development
of the internet. The World Wide Web, invented by and for particle
physicists, is for sharing information. The (computational) grid is
for the automated sharing resources on demand (such as computer
cycles, data storage, data, etc) in the same way as the electricity
grid gathers and distributes electrical power to consumers.  In this
analogy, the ILDG is intended to provide the sort of inter-grid link
that connects the UK and French power grids.  In the past, lattice
data has often be stored somewhat haphazardly in large archival
systems, without any systematic way of noting what or where the data
is.  This can lead to the situation where the data storage pattern is
\lq write-once, read-never\rq{}. 
There is a need to build separate, but interoperable,
systems which themselves, rather than the user, take care of recording
what the data are and where it resides

The assumption is that each collaboration participating in ILDG will
have its own data storage system.  Rather than impose some system on
all the collaborations, what is required is a uniform interface for
each grid to interact with another. This is sometimes called a
grid-of-grids, although technically this is a slightly misleading
term. However, what is important is that the owner of each resource
controls access to it.

Security of data and systems is obviously of paramount importance. The
identity of users is ensured by the use of X509 certificates and SSL
public/private key encryption.  There is a single login to the
grid. Time limited proxy certificates are generated for remote
operations. At a remote site, this proxy certificate is mapped to an
account, which could be a specific user account, or a non-specific
one. Certificates, for both machines and people, are issued by a
Certificate Authority (CA). The resource owner decides which CAs are
trusted and whether to accept their certificates. For instance, in the
UK the national e-science CA requires authentication from your home
institution and photographic ID before issuing a certificate. UK
citizens are not required to carry ID, so in the UK your virtual
identity is stronger than your real one!  ILDG will need to establish
a trust network of other CAs or could act as its own CA.

\subsection{Filenames and namespaces}
A namespace is collection of names which form a mathematical set. A
name can be given to the namespace, and this is often done using a URI
(Uniform Resource Identifier). A name of an object is then valid only
in that namespace. An XML namespace differs slightly in that it must
have a URI reference and can have internal structure in the
namespace. Both types of namespaces can be amalgamated. For example,
in the QCDML schema there is an element ``field'', this has a meaning
in the namespace of QCDML. In an application the name ``field'' may
already have a meaning, in which case there would be a conflict.  Both
namespaces can be made explicit, then ``app:field'' and
``qcdml:field'' are distinguishable.  

The Logical Filename (LFN) is
the name of a file in a particular namespace.  The namespace may
encapsulate different machines in different domains. The LFN is
not the physical address of the file. It is a name which uniquely
identifies that file in that namespace. Often it is a URI.

In a grid, the Replica Catalogue (RC) maps the LFN to the physical
address of the file. In a data grid, where there may be more than one
copy of the file, the RC tracks the number and location of the file
instances. Data access is via the RC and the LFN. The user doesn't
need to know anything about the file's physical location.

\subsection{QCDgrid - an example}
As an example of a data grid, and to highlight some of the ideas
above, we describe UKQCD's data grid, QCDgrid. QCDgrid is a member of
GridPP~\cite{gridpp} a collaboration of all UK groups in particle
physics working on the grid.  The hardware infrastructure is Linux PCs
(currently) running RH7.x with $\sim 1$ Tbytes RAID disk arrays. This
is relatively cheap and the RAID adds built in redundancy. Globus is
used for the low level middleware. The European Data Grid (EDG)
software will be used when and where possible, but at the moment the
software consists of custom written client tools and the central
control thread (CT) which runs the grid. Each of UKQCD's gauge
configurations has an accompanying XML metadata file.  Access to the
data is via the metadata. This can be searched in a browser which
queries the XML database (eXist, running in Tomcat).

The CT runs in a cycle on one node of the grid. As this is a single
point of failure, the node broadcasts its configuration files so that
if it goes down, another node can run the CT. The CT controls the flow
of data between users and storage nodes. The CT checks how much disk
space is available on each storage node, decides on which nodes data
is stored and from which node data is supplied to users, as well as
registering any changes with the RC. Data access is either with
command line tools and the LFN or via the metadata browser. The
browser GUI builds an Xpath query on the metadata, and the data can
then be downloaded. All UKQCD gauge configurations are stored and
accessed in this way.

\subsection{Grid of grids}
To implement the grid-of-grids we need to aggregate the RCs of
different grids. This has to be done in a secure fashion, such that
the owner controls the resource, and to distinguish between public,
restricted and private access. As a first step we make two RCs
simultaneously readable, and this requires a common interface between
the grids. Storage Resource Manager (SRM) seems ideally suited to this
role.  SRM is defined by the interface it presents to the outside
world, but behind that interface anything is allowed. The SRM standard
is still evolving. As a first step we aggregated the RCs of QCDgrid
and JLab running SRM as an interface. Version 1 of the specification
is employed, specifically a SOAP based XML protocol and the JLab Java
browser as a GUI for web services. This can list the LFN of the
QCDgrid and JLab RCs simultaneously

In the next step we would need to set up a trust network
of CAs, increase the functionality of the SRM, and interact with
the metadata. Both the QCDgrid software and the Globus SRM API are 
available under a GPL licence~\cite{nesc-sourceforge}.

\section{METADATA}
\label{sec:metadat}

QCD Markup Language (QCDML) is an XML schema for QCD lattice gauge
theory data. In this section, we summarise only the basic features of
the initial QCDML draft~\cite{ref:QCDMLdraft} for the description of
gauge configurations completed recently by the MDWG.  We focus here on
the description of physical parameters and the algorithm and on a
proposal for the binary format.

Collaborations submitting configurations to the ILDG will write a
QCDML XML document for each configuration and store it on their
database (and RC).  Researchers then issue a search query to the
database or RC and get information on availability. The database or RC
delivers the LFN in the ILDG namespace which then enables researchers
to retrieve the configuration using the ILDG middleware.

\subsection{Strategy and structure}
The MDWG has worked out the following strategy:
1) QCDML defines a minimal set of configuration information 
which researchers are usually interested in, without spoiling
extensibility to other lattice data.
2) In order to provide a unique description of information, 
we use principally ``element'' instead of ``attribute'' or
``value'', because one can define an allowed set of elements.
A module for each lattice action/field is prepared together with 
its precise definition written in a human readable glossary document 
and put on the ILDG web page. 

Each QCDML document consists of ``management'', 
``implementation'' (machine/code) and ``markov\_step'' sections.
Elements for a particular configuration such as trajectory
number are placed just under the ``markov\_step'' section. 
Ensemble information described in ``physics'', ``algorithm'' 
and ``precision'' parts are put together in ``markov\_chain'' 
subsection of the ``markov\_step''.

\subsection{Physics section}
\begin{figure}[t]
\includegraphics[scale=1.0]{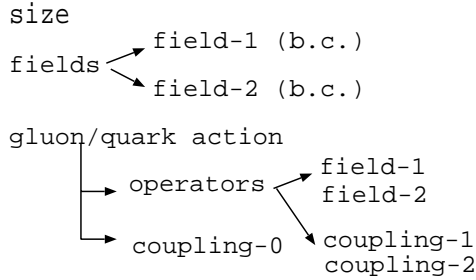}
\caption{Schematic diagram of the physics part.}
\label{fig:phys}
\vspace{-6mm}
\end{figure}

The physics part has a rich structure. 
A lattice action is divided into gluon and quark parts, each of 
which is a sum of operators (e.g. Wilson-Dirac operator) 
which in turn consist of fields 
(e.g. Wilson quark and link variables) and coupling parameters 
(e.g. $\kappa$).
A lattice field is in general independent of the lattice action.
Boundary conditions are a property of fields. 
Lattice size is also independent from field and action.
The structure then looks like Fig.~\ref{fig:phys} which
we have tried to describe straightforwardly in QCDML.

The number of dynamical quarks is the most important information
and is recorded in the quark action section, because different 
actions may be used for different flavours.
As an example, we record a configuration for $N_f=2+1$ QCD as
\begin{verbatim}
<quark>
  <n_sea_quarks>3</n_sea_quarks>
  <sw_quark_action>
    <n_quarks>2</n_quarks>
      ......
  </sw_quark_action>
  <sw_quark_action>
    <n_quarks>1</n_quarks>
      ......
  </sw_quark_action>
</quark>
\end{verbatim}

For a better understanding of the structure of the action part, 
we reproduce a complete pseudo-XML document below 
for the clover quark action. See Ref.~\cite{ref:QCDMLdraft} for 
details and further explanations.

\begin{verbatim}
<sw_quark_action>
  <n_quarks>2</n_quarks>
  <wilson_fermi_operator> 
    <field> link_gluon </field> 
    <field> wilson_quark </field> 
    <coupling>  
      <kappa> 0.1350 </kappa> 
    </coupling>  
  </wilson_fermi_operator> 
  <sw_fermi_operator> 
    <field> link_gluon </field> 
    <field> wilson_quark </field> 
    <coupling>  
      <c_sw> 2.02 </c_sw> 
      <determination> NP </determination> 
    </coupling>  
  </sw_fermi_operator> 
</sw_quark_action>
\end{verbatim}

\subsection{Algorithm section}
Since there are often many variants of a particular algorithm, the
MDWG came to the conclusion that it is not practical to describe an
algorithm precisely on a unique common basis. We therefore devolve
responsibility for the full description of an algorithm to the
supplier collaboration.

The QCDML defines only four common elements, ``name'', ``reference''
``exact'', and ``parameters''. In the ``reference'', we quote a paper
containing the algorithm when this is published. Otherwise, each
collaboration writes a glossary document, puts it on the web page, and
quotes the URL. Under the ``parameters'' element, each collaboration
places several elements whose names depend on collaboration and
algorithm. The collaboration name prefix (e.g. groupA:) is placed at
the beginning of each element. This (namespace) feature allows
different collaborations to continue to use their own terminology. The
working group asks all contributors to list at least the most basic
and important parameters.

\subsection{Binary format and distribution}
The MDWG also proposes a standard binary format of configurations as
an abstract/reference format. Each contributor will prepare a
C library to read their configuration (whose format can be different
from the standard one) into the standard format. Then the user can
convert it to their own format by writing a corresponding small C
program.

Another method of format conversion proposed depends on
BinX~\cite{ref:BinX} (an XML schema to describe binary format)
technology. Using a conversion tool which will be provided by the BinX
project and the MDWG, one can convert contributor format to user
format without referring to the standard format. For that purpose, we
ask contributors to prepare a BinX description of their own binary
format.

Details of the standard format are given in
Ref.~\cite{ref:QCDMLdraft}. An important point to note here is that we
propose to store only the first 2 rows of the 3x3 unitary
matrix. Users will reconstruct the third row using the unitarity
condition.

For the ILDG, keeping identification of configurations is important.
Information such as collaboration name and physics parameters is 
not defined in the standard format. Such information is recorded in
a corresponding QCDML document. 
Therefore, we propose to encapsulate the binary configuration,
QCDML document and BinX document into one file using DIME
technology, and distribute it via the ILDG.
The details for this will be discussed with the MWWG.

\section{PROGRESS AND FEEDBACK}
\label{sec:demo}
The QCDML proposal for gauge configuration metadata described above is
now open for comment and feedback from the community.  Please submit
your opinion to the MDWG via the forum page of the conference website
or write to them directly (qcdml@rccp.tsukuba.ac.jp).  Following this
feedback, the ILDG will adopt a modified version of this schema and
recommend that participants use it to mark up configurations which
they intend to share. The assumption is that normal etiquette on
collaboration, cooperation and publication will be observed when
making use of any \lq published\rq{} configurations.  It is expected
that the MDWG will then turn its attention to expansion of the schema
to cope with other lattice data objects (e.g. correlators) which
collaborations may wish to share.

\begin{figure}[ht]
\includegraphics[scale=0.45,angle=0]{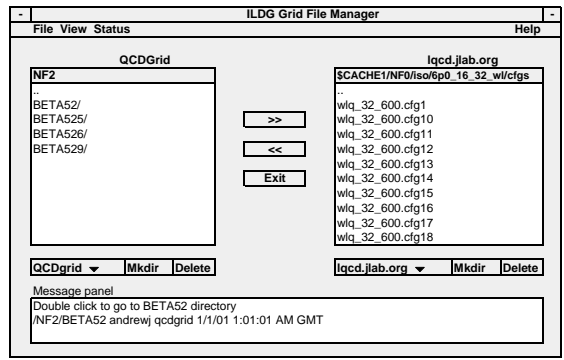}
\vspace{-0.5cm}
\caption{The prototype ILDG file manager used to browse remote data in QCDgrid (UK) and 
JLAB (USA).}
\label{fig:fbrowser}
\vspace{-0.5cm}
\end{figure}

The demonstration of intercontinental browsing of configuration
collections (initial goal for the MWWG) was partially successful --
within the limitations of bandwidth available at the conference site.
Collections in the UK (QCDgrid) and in the USA (JLab) were browsed
during the conference (see Fig~\ref{fig:fbrowser}).

As noted above, further work remains to be done to implement suitable
grid certification procedures.  The requirement here is to allow,
within a global grid context, easy read access to all potential users
and controlled write access to specified participants who themselves
may be geographically distributed. The MWWG also welcomes community
input on these and other issues.

A brief discussion session which followed the ILDG presentation
and demonstration indicated considerable potential interest in the
community. In particular, the desire was expressed for
better mechanisms for promoting technical feedback from lattice
collaborations to the working groups and for ratifying ILDG proposed
standards. It was also suggested that a third working group be
established to develop data access policies. These issues will be
considered at future ILDG workshops~\cite{ref:QCDMLdraft}.

\end{document}